\documentclass[a4paper,10pt]{article}

	\usepackage{enumerate,enumitem,hyperref,siunitx}
	\usepackage{bm,amsmath,amssymb,amsthm,physics}
	\usepackage{cite}
	\usepackage{graphicx}

\usepackage[font=small,labelfont=bf,figurename=Figure,labelsep=period]{caption}
	\usepackage[margin=1in]{geometry}
	\usepackage{float,pdflscape,authblk,setspace,lineno}
    \usepackage{combelow}
	\usepackage{algorithm,algpseudocode}
	\newtheorem{definition}{Definition}[section] 
	\usepackage{xcolor}

	\usepackage[capitalize]{cleveref}

\begin{document}

	\title{Exact identifiability analysis for a class of partially observed near-linear stochastic differential equation models}


	\author[1,2]{Alexander P Browning}
	\author[3]{Michael J Chappell}
	\author[2,4,5]{Hamid Rahkooy}
	\author[2]{\\Torkel E Loman}
	\author[2]{Ruth E Baker}
	\affil[1]{School of Mathematics and Statistics, University of Melbourne, Australia}
	\affil[2]{Mathematical Institute, University of Oxford, Oxford, OX2 6GG, United Kingdom}
	\affil[3]{School of Engineering, University of Warwick, Coventry, CV4 7AL, United Kingdom}
	\affil[4]{Max Planck Institute of Molecular Cell Biology and Genetics, Dresden, Germany}
	\affil[5]{Department of Mathematics, FIZ Karlsruhe - Leibnitz Institute for Information Infrastructure, Berlin, Germany}
		

\date{\today}
\maketitle
\footnotetext[1]{Corresponding author: apbrowning@unimelb.edu.au}


	\begin{abstract}
		\noindent Stochasticity plays a key role in many biological systems, necessitating the calibration of stochastic mathematical models to interpret associated data.  For model parameters to be estimated reliably, it is typically the case that they must be structurally identifiable. Yet, while theory underlying structural identifiability analysis for deterministic differential equation models is highly developed, there are currently no tools for the general assessment of stochastic models. In this work, we present a differential algebra-based framework for the structural identifiability analysis of linear and a class of near-linear partially observed stochastic differential equation (SDE) models. Our framework is based on a deterministic recurrence relation that describes the dynamics of the statistical moments of the system of SDEs. From this relation, we iteratively form a series of necessarily satisfied equations involving only the observed moments, from which we are able to establish structurally identifiable parameter combinations. We demonstrate our framework for a suite of linear (two- and $n$-dimensional) and non-linear (two-dimensional) models. Most importantly, we define the notion of structural identifiability for SDE models and establish the effect of the initial condition on identifiability. We conclude with a discussion on the applicability and limitations of our approach, and potential future research directions in this understudied area.
	\end{abstract}
	
	
\clearpage
\section{Introduction}

Many biological systems are either intrinsically noisy \cite{Abkowitz:1996,Elowitz.2000,Bruckner.2020} or subject to noisy external forces \cite{Turelli:1977,Anderson.2007,Hidalgo.2015,Browning.2024c1p,Browning.202115}, motivating the application of stochastic mathematical models to interpret any associated data \cite{Wilkinson:2012,Wilkinson:2009}. In particular, information about the temporal dynamics of systems observed at statistical equilibrium is often not accessible through deterministic models, but rather encoded in higher-order statistics such as the autocovariance of temporally resolved observations \cite{Munsky:2009,Komorowski:2011}. Even within non-equilibrium systems, stochastic models may be able to extract more information about model parameters than their deterministic counterparts \cite{Bressloff:2017,Browning.2020,Huynh.2023}. 

Tools that enable the calibration of stochastic models---and even those that allow for model discovery \cite{Nabeel.2022}---are now widely available \cite{Golightly:2011,Frohlich:2016,Schnoerr:2017,Warne:2019w3q}. For stochastic differential equation (SDE) models, methods fall into two general categories: approximate methods, which include among others the linear noise approximation \cite{Komorowski:2009}, the finite state projection \cite{Fox:2016}, and approximate Bayesian computation \cite{Picchini:2012}; and exact methods \cite{Kalogeropoulos:2011}, typically based on a particle filter approach to the exact likelihood \cite{Beaumont:2003,Andrieu:2010,Warne:2020}. While many of these methods allow for uncertainty quantification---through either a posterior distribution or parameter confidence intervals \cite{Warne:2020}---fundamental theoretical questions relating to whether model parameters are \textit{identifiable} remain unanswered. Specifically, for a parameter to be \textit{practically identifiable} from a given set of experimental data it should first and necessarily be \textit{structurally identifiable}: a mathematical property of the model which establishes whether the map from model parameters to model outputs (typically a subset or linear combination of model states that are observed) is injective \cite{Bellman.1970,Walter:1987}. This question of structural identifiability amounts to asking whether a parameter can be estimated in the limit of infinite, noise-free, data.

Theory underlying the structural identifiability of ordinary differential equation (ODE) systems is now highly developed \cite{Chis:2011}. Methods developed out of differential algebra (DA)\cite{Ritt.1950,Ljung:1994} involve the algorithmic formulation of a set of \textit{input-output equations}, a set of necessarily satisfied polynomial equations that depend solely on model parameters, observed system states, and derivatives thereof \cite{Margaria.2001,Bellu:2007,Renardy.2022}. These input-output equations provide a map from the parameters (the inputs) to the the set of observations (the outputs). These relations are also analogous to representing a system of first-order ODEs as a system of higher-order ODEs involving only a subset of state variables: in the case of identifiability, involving only state variables that are observed. Necessarily, these equations always involve higher order derivatives, complicating direct extension to stochastic systems that are not differentiable. By ensuring that these polynomials are in some sense monic and hence unique, and if linear independence of the monomial terms can be shown or guaranteed, structurally identifiable parameter combinations can be read off as the set of parameter-dependent coefficients \cite{Renardy.2022}. 

Our more recent work \cite{Browning.2024d0n}, among others \cite{Renardy.2022,Ciocanel.2024,Byrne.2025}, extends the DA approach to a class of near-linear partial differential equation systems, for which the same methodology broadly applies. Systems of SDEs, on the other hand, cannot be studied in the same way. We partially address this in our previous work \cite{Browning.2020} by studying identifiability through an ODE system that describes the time-evolution of the statistical moments of the system of SDEs. This approach is both exact and exhaustive for linear systems of SDEs. For non-linear systems, which are not characterised by a finite number of statistical moments, we resorted to studying an approximate system by applying a series of moment closure approximations. While we demonstrate that the approximate approach can provide useful insight, it remains unclear how well structural identifiability results for the approximate system translate to the exact system.  Also ambiguous are the observation regimes and initial conditions under which results from study of the statistical moments apply to individual trajectories.

In this work, we present a DA-based framework for an exact structural identifiability analysis of partially observed linear as well as a class of non-linear systems of SDEs.  We first, in \cref{sec2}, provide a formal definition of structural identifiability for both deterministic---for which we present a brief didactic example---and, for the first time, stochastic models. Specifically, we define what it means for a stochastic model to be structurally identifiable given a hypothetical joint distribution of model outputs across arbitrary sets of time points (i.e., from observations of sample paths). 

Given the ubiquity of linear systems throughout both biology and engineering, we begin with a complete analysis of structural identifiability in a general two-dimensional linear Gaussian (i.e., Ornstein-Uhlenbeck) system (\Cref{sec:2d_ou}). This system is unique in that it possesses an analytical solution, and in that the behaviour is completely characterised by the dynamics of the first two statistical moments. In the case of discrete-time observations, structural non-identifiability could be detected locally through the Fisher information matrix \cite{Khasminskii.1999,Raue:2009}. The tractability of this simple system allows us to demonstrate how results constructed through the moment approach \cite{Browning.2020} relate to observations of the SDE at statistical equilibrium: the set of \textit{observed quantities} correspond to an initial condition where the unobserved states are constrained statistically to the stationary distribution conditional to the observed states. We then extend this analysis to provide a non-exhaustive set of structurally identifiable parameter combinations in a partially observed $n$-dimensional linear Gaussian (i.e., Ornstein-Uhlenbeck) system (\Cref{sec:nd_ou}).

The final linear system we study is an analogous two-dimensional system subject to geometric noise (\Cref{sec:geo}). This system is not Gaussian, and does not, in general, possess a closed-form analytical solution. Being linear, the moment approach can still be applied directly (and exactly) using our previous methodology. However, the system is only completely characterised by the infinite system of moments: we do not expect to be able to establish an exhaustive set of structurally identifiable parameters by studying a closed second-order system. Our contribution, in this work, is to proceed by establishing a recurrence relation for the dynamics of each statistical moment. From this, we iteratively construct a series of input-output equations---as we term, ``necessarily satisfied equations''---that yield a (possibly non-exhaustive) set of identifiable parameter combinations.

Importantly, our iterative approach does not require linearity, and is applicable to a broader class of SDE models that cannot be analysed exactly at all using existing approaches. We proceed to demonstrate the applicability of our framework to a class of two-dimensional non-linear SDE systems; specifically, systems that remain linear in the unobserved variable, and systems where interaction-like terms (i.e., the product of unobserved and observed variables) do not appear in the governing equations for variables that are observed. In particular, we demonstrate our approach through analysis of a stochastic logistic model, a simplified Lotka-Volterra model, and a chemical Langevin equation derived for a second-order system of chemical reactions. We conclude with a discussion on the practical application of our framework, its limitations, and future research directions. Code used to perform all analyses are available as supporting material on GitHub\footnote{See \href{https://github.com/ap-browning/sde_structural_identifiability}{github.com/ap-browning/sde\_structural\_identifiability}}.

\section{Preliminaries}\label{sec2}

For deterministic models, the notion of structural identifiability is well established \cite{Evans.2013,Renardy.2022}, and defined as follows.
\begin{definition}\label{def_ode}
	A deterministic model, $\mathbf{m}(\bm\theta,t)$, that maps parameters, $\bm\theta$, to observed model outputs, is said to be \textnormal{globally structurally identifiable} if and only if $\mathbf{m}(\bm\theta,t) = \mathbf{m}(\bm\theta^*,t)$ for all $t \in \mathbb{R}$ and for almost all initial conditions implies that $\bm\theta = \bm\theta^*$. If the implication also holds for a function or combination of parameters $\varphi(\bm\theta) = \varphi(\bm\theta^*)$, then $\varphi(\cdot)$ is said to be globally structurally identifiable. Finally, if the implication holds for $\bm\theta^* \in \mathcal{B}(\bm\theta)$, where $\mathcal{B}(\bm\theta)$ denotes an arbitrarily small region centred at $\bm\theta$, then the model or combination of parameters is said to be \textnormal{locally structurally identifiable}. If a parameter or combination of parameters is not globally or locally structurally identifiable, then we say that is non-identifiable.
\end{definition}

For example, consider the ODE model $x'(t) = (a - b) x(t)$ subject to $x(0) = x_0$ and for $a \neq b$. With one state variable, the model must be fully observed, such that $m(\bm\theta,t) = x(t)$ for parameter vector $\bm\theta = \big(a,b,x_0\big)^\intercal$. The exact solution is given by $x(t) = x_0 \exp((a - b) t)$, from which it can be seen that the difference $\varphi(\bm\theta) = a - b$ and the parameter $x_0$ must both be preserved for the model solution to remain unchanged. Thus, the model is not globally structurally identifiable, but the reparameterised model $x'(t) = c x(t)$ with $c = a - b$ is globally structurally identifiable.

For stochastic models, we interpret the map as being from model parameters to the hypothetical \textit{distribution} of model outputs. Denoting the hypothetical joint probability density function of model outputs at a set of time points $\{t_i\}_{i=1}^N$ as $\pi(\bm{x}(t_1), \dots, \bm{x}(t_N); t_1, \dots, t_N, \bm\theta)$, where $\bm{x}(t) \in \mathbb{R}^m$ corresponds to the observed model output at time $t$, we can expand \Cref{def_ode} to stochastic models. 
\begin{definition}\label{def_sde}
	A stochastic model, the observed outputs of which are characterised by the probability density function $\pi(\bm{x}(t_1), \dots, \bm{x}(t_N); t_1,\dots,t_N,  \bm\theta)$, is said to be \textnormal{globally structurally identifiable} if and only if $$\pi(\bm{x}(t_1), \dots, \bm{x}(t_N); t_1,\dots,t_N, \bm\theta) = \pi(\bm{x}(t_1), \dots, \bm{x}(t_N); t_1,\dots,t_N, \bm\theta^*)$$ for all $\{t_i\}_{i=1}^N \in \mathbb{R}^N$, for all $N$, implies that $\bm\theta = \bm\theta^*$. If the implication holds for a scalar function or combination of parameters $\varphi(\bm\theta) = \varphi(\bm\theta^*)$, the combination $\varphi(\cdot)$ is said to be \textnormal{globally structurally identifiable}. Finally, if the implication holds for $\bm\theta^* \in \mathcal{B}(\bm\theta)$, where $\mathcal{B}(\bm\theta)$ denotes an arbitrarily small region centred at $\bm\theta$, then the model or combination of parameters is said to be \textnormal{locally structurally identifiable}.
\end{definition}
In many cases, the output distribution may not include the full system state; for example, in systems of chemical reactions where only a subset of species is observed \cite{Raue:2009}. Furthermore, we that we may have that $\varphi : \bm\theta \mapsto \theta_i$ for $\theta_i \in \bm\theta$, in which case the individual parameter $\theta_i$ is said to be structurally identifiable. Structural identifiability analysis is primarily concerned with establishing (a potentially exhaustive) set of these structurally identifiable parameters and parameter combinations \cite{Renardy.2022}.

\section{Partially observed linear systems}\label{seclinear}

We begin with an exhaustive analysis of linear systems of SDEs, for which the system of ODEs describing the statistical moments are closed at every order \cite{Browning.2020}. Consequentially, for linear systems, the first moment (i.e., the mean) depends only on the drift, and not the diffusive component of the system. Furthermore, linear models with state-independent diffusion, such as Ornstein-Uhlenbeck processes (Sections \ref{sec:2d_ou} and \ref{sec:nd_ou}), are Gaussian processes, such that their behaviour is completely characterised by a closed system describing the first two statistical moments. For linear models with state-dependent diffusion (\Cref{sec:geo}), the moment equations are closed at every order, but the infinite system is required to fully characterise the process.

	\subsection{Two-state Ornstein-Uhlenbeck process}\label{sec:2d_ou}
	
	We first consider a general two-state Ornstein-Uhlenbeck model \cite{Gardiner}, given in factorised form as
		\begin{equation}\label{sde1}
			\dd \mathbf{x}(t) = -\mathbf{A}\left(\mathbf{x}(t) - \mathbf{b}\right) \dd t + \mathbf{S} \,\dd\mathbf{W},\\
		\end{equation}
	for $\mathbf{x}(t) = [x(t),y(t)]^\intercal$, where $x(t)$ is the observed state, $y(t)$ is unobserved, and $\mathbf{W} \in \mathbb{R}^2$ is a two-dimensional Wiener process with independent components. We are interested in the identifiability of the unknown coefficient matrices and vectors $\mathbf{A}$, $\mathbf{b}$, and $\mathbf{S}$, where we set
		\begin{equation}\label{linear_mat_pars}
			\mathbf{A} = \begin{pmatrix} a & b \\ c & d \end{pmatrix},\quad 
			\mathbf{b} = \begin{pmatrix} e \\ f \end{pmatrix},\quad
			\mathbf{S} = \begin{pmatrix} p & q \\ r & s \end{pmatrix}.
		\end{equation}

	While much of the proceeding mathematical analysis does not explicitly require it, we assume that all eigenvalues of $\mathbf{A}$ are positive, such that a stationary distribution for $\mathbf{x}(t)$ exists. This allows us to consider regimes in which trajectories of \cref{sde1} are sampled at statistical equilibrium. Additionally, we set $q = 0$ to avoid the trivial structural non-identifiability of $\mathbf{S}$ that arises from over-parameterisation of the Gaussian random variable $\mathbf{S} \,\mathbf{W}$, which is characterised by three effective parameters governing  the variance and correlation associated with each of the two state variables. 
	
	As a Gaussian process, the behaviour of \cref{sde1} is characterised entirely by its first two moments. We denote by
		\begin{equation}
			m_{i,j}(t) = \langle x^i(t) y^j(t) \rangle,
		\end{equation}
	the $(i,j)$-th raw moment, where $\langle \cdot \rangle$ denotes an expectation. Applying It\^{o}'s lemma \cite{Socha:2008}, the system of governing moment equations is given by
	\begin{equation}\label{sde1_moments}
	\begin{aligned}
		m_{1,0}'(t) 	&= a e + b f - a m_{1,0}(t) - b m_{0,1}(t),\\
		m_{0,1}'(t) 	&= c e + d f - c m_{1,0}(t) - d m_{0,1}(t),\\
		m_{2,0}'(t) 	&= 2\Big[(a e + b f)m_{1,0}(t) - a m_{2,0}(t) - b m_{1,1}(t)\Big] + p^2,\\
		m_{0,2}'(t) 	&= 2\Big[(c e + d f)m_{0,1}(t) - d m_{0,2}(t) - c m_{1,1}(t)\Big] + r^2 + s^2,\\
		m_{1,1}'(t)  &= (c e + d f) m_{1,0}(t) +  (a e + b f) m_{0,1}(t) - c m_{2,0}(t) - b m_{0,2}(t) - (a + d) m_{1,1}(t) + p r.
	\end{aligned}
	\end{equation}

	Following our previous work \cite{Browning.2020}, we can now assess structural identifiability directly from \cref{sde1_moments} by considering that only moments associated with the marginal distribution of the observed variable $x(t)$ (i.e., $m_{1,0}(t)$ and $m_{2,0}(t)$) are observed. Applying the DA-based \textit{StructuralIdentifiability.jl} package \cite{Dong.2023} (supplementary code), we obtain the following set of structurally identifiable parameter combinations
		\begin{equation}\label{ptilde1}
			\tilde{\mathcal{P}} = \{a+d,ad - bc,e\},
		\end{equation}
	in addition to a set of observable quantities that depend on both the parameters and the system state (i.e., the initial condition). In the case that $ad - b c \neq 0$, which follows from our assumption that all eigenvalues of $\mathbf{A}$ are positive, we can simplify this set of observed quantities to a set of intermediate quantities that are \textit{observable}:
		\begin{equation}\label{sde1_oq}
		\begin{aligned}
			\mathcal{Q}_1 &= m_{1,0}(t), \\
			\mathcal{Q}_2 &= m_{2,0}(t),\\
			\mathcal{Q}_3 &= d m_{1,0}(t) - b m_{0,1}(t) + b f - d e,\\
			\mathcal{Q}_4 &= a \Big(m_{1,0}^2(t) - m_{2,0}(t)\Big) - b\Big(m_{1,1}(t) - m_{1,0}(t)m_{0,1}(t)\Big) + \frac{p^2}{2},\\
			\mathcal{Q}_5 &=  c\Big(m_{1,0}^2(t) - m_{2,0}(t)\Big) + (a - d) b \Big(m_{1,1}(t) - m_{1,0}(t)m_{0,1}(t)\Big)\\
			&\qquad\quad - b^2\Big(m_{0,1}^2(t) - m_{0,2}(t)\Big) + p (d p - b r),\\
			\mathcal{Q}_6 &= \left(a + \frac{7d}{5}\right) m_{1,0}^2(t) + \dfrac{2b}{5}\Big(m_{1,1}(t) -  m_{1,0}(t)m_{0,1}(t)\Big)\\&\qquad\quad  + \frac{3a + d}{5}m_{2,0}(t) - \dfrac{(d p - b r)^2 + b^2 s^2}{10(ad - b c)} + \dfrac{3p^2}{10}.
		\end{aligned}
		\end{equation}
	As with identifiability, an observable quantity is one that must be preserved for the observed model outputs to remain unchanged. While this is also true for quantities that are identifiable (see \cref{def_sde}), an observable quantity defines an algebraic relationship between model states \cite{Diop.1991}. For example, from \cref{ptilde1} we see that $b$ and $c$ can be chosen arbitrarily, subject to the constraint that the identifiable quantity $ad - bc$ remains unchanged (this will necessarily be true in the case that the product $bc$ is held constant). Similarly, the initial conditions $m_{0,1}(0)$, $m_{0,2}(0)$, and $m_{1,1}(0)$ can be chosen arbitrarily, provided that the observed quantities in \cref{sde1_oq} remain unchanged. Before proceeding, we note that \textit{all} parameters are identifiable when both states are observed (supplementary code).
		
	\begin{figure}[!t]
		\centering
		\includegraphics[scale=1]{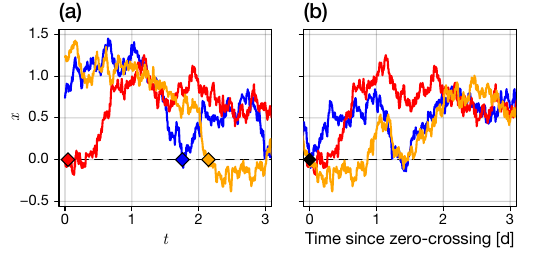}
		\caption{\textbf{Translated trajectories.} For autonomous SDEs, we translate time such that trajectories sampled from their stationary distribution cross the axis at $t = 0$. While moments of the unobserved quantity $y(t)$ are not observed, they correspond to the known conditional stationary distribution $y(0) | x(0) = x_0$. }
		\label{fig1}
	\end{figure}

\begin{figure}[!t]
		\centering
		\includegraphics[width=\textwidth]{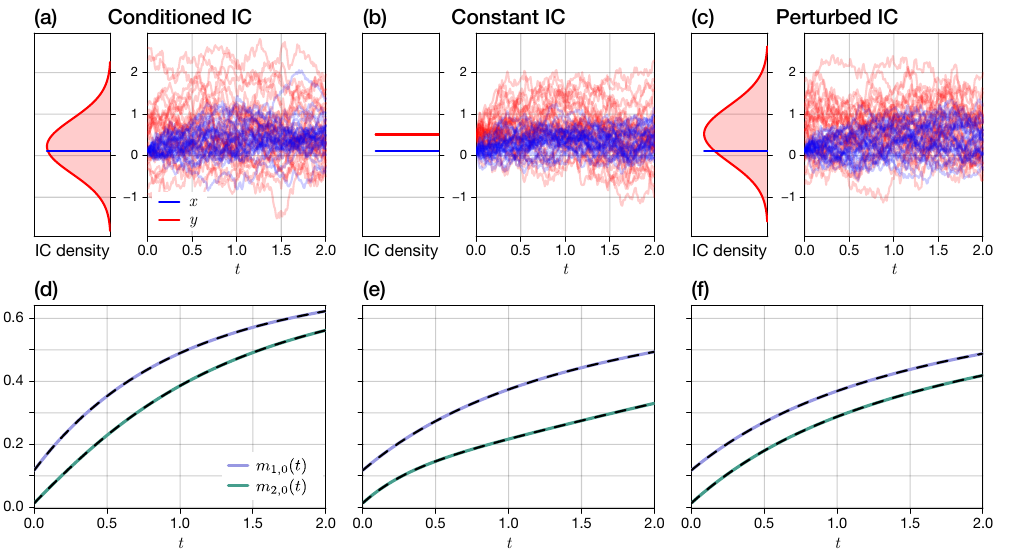}
		\caption{\textbf{Initial condition and non-identifiability in the Ornstein-Uhlenbeck model.} (a--c) Simulated data sets showing the observed variable $x(t)$ (blue) and the unobserved variable $y(t)$ (red) for various initial conditions. The density function for the initial condition is shown to the left of each set of realisations. (d--f) The observed raw moments $m_{1,0}(t)$ (purple; corresponding to the mean of $x(t)$) and $m_{2,0}(t)$ (turquoise; corresponding to the variance of $x(t)$ through $\mathrm{Var}(x(t)) = m_{2,0}(t) - m_{1,0}^2(t)$). Parameters are randomly chosen and are fixed between initial conditions. We also show (black dashed) the solutions for a second, distinct, randomly chosen set of parameters with the same set of identifiable parameter combinations.}
		\label{fig2}
	\end{figure}

	We interpret structural identifiability for SDEs in the distributional sense (see Definition 2). In the context of data, the question of structural identifiability effectively considers a limiting regime in which an arbitrarily large number of trajectories are observed over a arbitrarily long period of time such that the \textit{distribution} of model outputs, between all time points, is observed.  As the system is autonomous, we can, without loss of generality, translate (in $t$) all observed trajectories such that $\mathbb{E}(x(0)) = x_0$ and $\mathrm{Var}(x(0)) = 0$. We demonstrate such a translation in \cref{fig1}. For example, consider the evolution of the moment equations in the case that trajectories are translated such that $x(0) = 0$, such that $m_{1,0}(0) = m_{2,0}(0) = m_{1,1}(0) = 0$ (although, any crossing point in $x(0)$ can be chosen such that results hold for systems that cross $x(t) = 0$ with very small probability). We can then substitute the initial condition at $t = 0$ into the set of observed quantities to find additional structurally identifiable parameter combinations. From $\mathcal{Q}_4$, all terms except the last vanish, indicating that $p^2$ and, therefore, $p$ is structurally identifiable. Another identifiable combination can also be drawn from $\mathcal{Q}_6$. After some simplification, the set of identifiable parameter combinations is then given by
		\begin{equation}\label{sde1_si}
			\mathcal{P} = \{a+d,ad - bc,e,p,(d p - b r)^2 + b^2 s^2\}.
		\end{equation}
	Thus, only the trace and determinant of $\mathbf{A}$ (relating to the eigenvalues of $A$, we address this observation in \Cref{sec:nd_ou}), and not its individual constituents, are identifiable. The remaining observed quantities $\mathcal{Q}_3$ and $\mathcal{Q}_5$, do not provide additional information about the parameters of this model. Rather, they define the relationship between the moments of the unobserved variable $y(t)$ and the observed variable $x(t)$. Specifically, these quantities will always be preserved should the distribution of the unobserved variable $y(t)$ follow the stationary distribution of the process conditioned on the observed value of $x(t)$ (supplementary Mathematica code). We therefore interpret the set of identifiable parameter combinations given in \cref{sde1_si} as those that can be established from correlated observations of a stationary SDE; i.e., experiments in which the experimentalist does not intervene. In \cref{fig2}a,d, we demonstrate how two distinct parameter sets can be chosen with quantities in \cref{sde1_si} preserved, such that the model observations are identical in distribution.

	\subsubsection*{Experimentally relevant initial conditions}
	
	Our analysis thus far corresponds to a scenario where the experimentalist does not intervene in a system at statistical equilibrium. We now consider two additional initial conditions that are experimentally relevant. First, a \textit{constant initial condition} where $y(0)$ is not observed, but where, conditional on $x(0) = x_0$, $\mathrm{Var}(y(0)) = 0$. Secondly, a \textit{perturbed initial condition} where we consider an experimentalist that intervenes initially, such that $y(0)$ is at its stationary distribution but where $x_0$ is (independently) fixed at a pre-determined value. In both scenarios, we expect the number of identifiable combinations only to increase, since the system will, effectively, return to the equilibrium state studied previously for $t \gg 0$. Furthermore, it is only from the observed quantities $\mathcal{Q}_3$ and $\mathcal{Q}_5$ (\cref{sde1_oq}) that we expect to gain additional information.
		
	\begin{enumerate}
		\item \textit{Constant initial condition.} Here, we assume that $\mathrm{Var}(y(0)) = 0$, such that $m_{0,2}(0) = m_{0,1}^2(0)$. This represents an experiment initiated, for example, with an unknown but fixed concentration of each chemical $x$ and $y$, but where only $x$ can be observed through time. Substituting $m_{0,2}(0) = m_{0,1}^2(0) = y_0^2$ (and similarly for $x_0$) into the observable $\mathcal{Q}_5$ yields $dp - b r$ as an additional structurally identifiable parameter combination, while $\mathcal{Q}_3$ yields an observed relationship between the observed and unknown initial conditions, $x_0$ and $y_0$, respectively. Thus, the quantities 
			\begin{equation*}
				\mathcal{P} = \{a+d,ad - bc,e,p,d p - b r, b^2 s^2,d (x_0 - e) + b(y_ 0 - f)\},
			\end{equation*}
		are now identifiable. We demonstrate structural non-identifiability for the constant initial condition in  \cref{fig2}b,e.
		
		\item \textit{Perturbed initial condition.} Here, we consider that the system is initially stationary, but that $x(0)$ is perturbed independently of $y(0)$. Thus, we assume that $m_{0,1}(0) = f$ and
			\begin{equation*}
				m_{0,2}(0) = \dfrac{c^2p^2+a(a+d)(r^2 + s^2) - c(2apr + b(r^2 + s^2)}{2(a+d)(ad - bc)} + f^2,\\
			\end{equation*}
		which is derived from the analytical solution to the stationary distribution of \cref{sde1} \cite{Meucci.2009} (supplementary code). From $\mathcal{Q}_3$ we find that that $d(x_0 - e)$ is identifiable, and thus $d$ and by extension $a$ are now identifiable. $\mathcal{Q}_5$ provides no new information. The set of identifiable parameters is now given by
		\begin{equation*}
			\mathcal{P} = \{a,bc,d,e,p,(d p - b r)^2 + b^2 s^2\}.
		\end{equation*}
		We demonstrate structural non-identifiability for the perturbed initial condition in \cref{fig2}c,f.

	\end{enumerate}

	\subsubsection*{Independent observations of both states}
	
	Another experimentally relevant scenario is one in which both species can be observed through separate \textit{independent} experiments; for example, in the case that only a single fluorescent marker is available. Mathematically, this implies that the marginal distribution of each state is observed (including each respective autocovariance), but the joint distribution for both $x(t)$ and $y(t)$ is not.	We stress that this cannot be assessed by considering analysis of the moment equations where all moments but $m_{1,1}(t)$ are observed, as underlying this approach is the problematic assumption of a shared initial condition. If the states are truly observed independently, this will \textit{not} be the case: rather, the correct initial condition to consider is one in which the unobserved variable follows its stationary distribution conditioned on the observed variable. This way, observations of each state are made independently, and the only assumption is that the system is observed at statistical equilibrium.	

	To make progress, we first repeat the earlier analysis on \cref{sde1} in the case that it is only $y(t)$ that is observed, such that
		\begin{equation}\label{sde1_si_y}
			\mathcal{P}_y = \{a+d,ad - bc,f,r^2 + s^2,(c p - a r)^2 + a^2 s^2\},
		\end{equation}
	is the set of identifiable parameter combinations. We can then combine \cref{sde1_si,sde1_si_y} and simplify to obtain the set of identifiable parameter combinations from independent observations of each state, yielding
		\begin{equation}\label{sde1_si_xy}
			\mathcal{P}_{xy} = \{a+d,ad - bc,e,f,p,r^2 + s^2,(c p - a r)^2 + a^2 s^2,(d p - b r)^2 + b^2 s^2\}.
		\end{equation}
	%
		
	\begin{figure}[!t]
		\centering
		\includegraphics[scale=1]{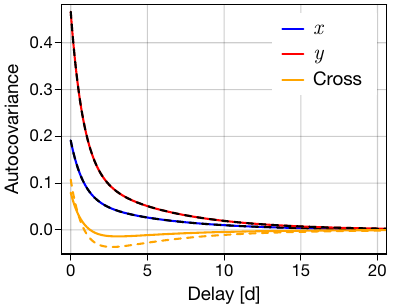}
		\caption{\textbf{Non-identifiability from independent observations of all states in the two-dimensional linear model.} We plot the autocovariance functions $\langle x(0), x(\tau) \rangle$, and similarly for $y(t)$, for two distinct parameter sets that preserve the structurally identifiable combinations given in \cref{sde1_si_xy}. The autocovariance function is shown in colour for the first parameter set (blue, for $x(t)$; red, for $y(t)$), and in black-dashed for the second. We also show that the cross-autocovariance function $\mathrm{Cov}(x(0),y(\tau))$ is distinct for each parameter set.}
		\label{fig3}
	\end{figure}
	
	Again, only quantities relating to the eigenvalues of $\mathbf{A}$ are identifiable, and not individual constituents. However, this observation regime is sufficient to constrain structural identifiability to a one-dimensional curve: we establish in \cref{sde1_si_xy} eight identifiable parameter combinations from a set of nine total parameters. In \cref{fig3}, we demonstrate how preserving these identifiable quantities is sufficient to yield distinct parameter sets that give rise to indistinguishable model outputs. As a stationary Gaussian process (with observed mean $[e,f]^\intercal$), the behaviour can be completely characterised by the autocovariance function. Thus, the results in \cref{fig3} show how these quantities preserve the marginal autocovariance functions, but not the cross-covariance function $\mathrm{Cov}(x(0),y(\tau))$, where $\tau$ denotes the delay, that characterises the joint distribution between $x(t)$ and $y(t)$, which is not observed.

	\subsection{$n$-dimensional Ornstein-Uhlenbeck process}\label{sec:nd_ou}
	
	We now consider that $\mathbf{x}(t)$ is an $n$-dimensional Ornstein-Uhlenbeck process, described by \cref{sde1} with $\mathbf{x}(t) = [\mathbf{x}_1^\intercal(t), \mathbf{x}_2^\intercal(t)]^\intercal$ where $\mathbf{x}_1(t) \in \mathbb{R}^m$ are the observed states, and $\mathbf{x}_2(t) \in \mathbb{R}^{n-m}$ are the unobserved states. As the $n$-dimensional system is still linear, the mean is described entirely with an ODE recovered from \cref{sde1} with $\mathbf{S} = \mathbf{0}$. Therefore, we expect that all parameters that are structurally identifiable in this ODE model will also be  identifiable from the SDE. What remains unclear is both the identifiability of $\mathbf{S}$, and how $\mathbf{A}$ and $\mathbf{S}$ interact to produce parameter combinations involving terms in $\mathbf{A}$ that are identifiable from the SDE model, but not identifiable from the ODE model.
	
	We partition the matrices $\mathbf{A}$, $\mathbf{b}$, and $\mathbf{S}$ as
		\begin{equation}
			\mathbf{A} = \begin{pmatrix} 
 				\mathbf{A}_{11} & \mathbf{A}_{12} \\
 				\mathbf{A}_{21} & \mathbf{A}_{22}
 			\end{pmatrix},\quad 
 			\mathbf{b} = \begin{pmatrix} 
 				\mathbf{b}_{1}\\
 				\mathbf{b}_{2}
 			\end{pmatrix},\quad 
 			\mathbf{S} = \begin{pmatrix} 
 				\mathbf{S}_{11} & \mathbf{S}_{12} \\
 				\mathbf{S}_{21} & \mathbf{S}_{22}
 			\end{pmatrix},\quad 
		\end{equation}
	where $\mathbf{A}_{11}, \mathbf{S}_{11} \in \mathbb{R}^{m \times m}$ and $\mathbf{b}_1 \in \mathbb{R}^{m}$. As the behaviour of the system is characterised by $\mathbf{S}\mathbf{S}^\intercal$ and not solely by $\mathbf{S}$, without loss of generality we assume that $\mathbf{S}$ is lower triangular, such that both $\mathbf{S}_{11}$ and $\mathbf{S}_{22}$ are also lower triangular and that $\mathbf{S}_{12} = \mathbf{0}$.
	
	As an Ornstein-Uhlenbeck process, $\mathbf{x}(t)$ follows a Markov Gaussian process. By extension, as a partition of $\mathbf{x}(t)$, $\mathbf{x}_1(t)$ also follows a Gaussian process, but is not Markovian. The behaviour of the observed states is, therefore, defined entirely in terms of the stationary mean and autocovariance function. This mean is given by $\mathbf{b}_1$, and thus we see immediately that all components of $\mathbf{b}_1$ are structurally identifiable. 

	We next consider an analytical expression for the autocovariance function of the entire state, given by \cite{Meucci.2009}
		\begin{equation}\label{ou_nd_fullcov}
			\mathrm{Cov}(\mathbf{x}(0),\mathbf{x}(t))= 
			\begin{pmatrix}
				\mathbf{\Sigma}_\infty & \mathbf{\Sigma}_\infty \mathrm{e}^{-\mathbf{A}^\intercal t}\\
				\mathrm{e}^{-\mathbf{A}t}\mathbf{\Sigma}_\infty & \mathbf{\Sigma}_\infty
			\end{pmatrix},
		\end{equation}
	where
		\begin{equation}\label{ou_nd_cov}
			\mathrm{vec}(\mathbf{\Sigma}_\infty) = (\mathbf{A} \oplus \mathbf{A})^{-1}\mathrm{vec}(\mathbf{S}\mathbf{S}^\intercal),
		\end{equation}
	defines $\bm\Sigma_\infty$, the covariance of $\mathbf{x}(t)$ at statistical equilibrium. Here, $\mathrm{vec}(\cdot)$ denotes the column-wise vector operator, and $\oplus$ the Kronecker sum. We are particularly interested in the autocovariance function for the observed states, which (along with the mean) completely defines the behaviour of the Gaussian process. We thus define
		\begin{equation}\label{rho}
			\bm\rho(t) := \Big[\mathrm{e}^{-\mathbf{A}t} \mathbf{\Sigma}_\infty\Big]_{11},
		\end{equation}
	where we denote by $[\mathbf{M}]_{11} = \mathbf{M}_{11} \in \mathbb{R}^{m \times m}$ the leading block of a matrix $\mathbf{M} \in \mathbb{R}^{n \times n}$. Structural non-identifiability corresponds, therefore, to two distinct sets of parameter values yielding equal $\mathbf{b}_{1}$ and equal $\bm\rho(t)$.

	We make progress toward establishing a set of structurally identifiable parameter combinations by considering the diagonalisation of the matrix $\textbf{A}$. We do not, however, expect this set to be exhaustive: additional parameter combinations may become identifiable from, as we soon demonstrate, consideration of the stationary covariance of the observed stochastic process. Assuming that all eigenvalues of $\mathbf{A}$, denoted by $\lambda_1,\ldots,\lambda_n$, are distinct---this is in general true for random matrices \cite{Edelman.2005}---we write
		\begin{equation}\label{rho_expanded}
			\bm\rho(t) = \Big[\mathbf{V} \mathrm{e}^{-\mathbf{D} t} \mathbf{V}^{-1} \mathbf{\Sigma}_\infty\Big]_{11} = \left[\sum_{i=1}^n\left(\mathrm{e}^{-\lambda_i t}\mathbf{v}_i\tilde{\mathbf{v}}_i^\intercal \bm\Sigma_\infty\right)\right]_{11} = \sum_{i=1}^n \left(\mathrm{e}^{-\lambda_i t}\Big[\mathbf{v}_i\tilde{\mathbf{v}}_i^\intercal \bm\Sigma_\infty\Big]_{11}\right),
		\end{equation}
	where $\mathbf{D} = \mathrm{diag}(\lambda_1,\ldots,\lambda_n)$, where $\mathbf{V} = [\mathbf{v}_1,\ldots,\mathbf{v}_n]$ is a matrix with columns corresponding to the respective eigenvectors, and where we denote $\mathbf{V}^{-1} = [\tilde{\mathbf{v}}_1,\ldots,\tilde{\mathbf{v}}_n]$. As we can write $\bm\rho(t)$ as the sum of exponentials, with exponents equal to the eigenvalues of $\mathbf{A}$, we expect that, in general, the eigenvalues of $\mathbf{A}$ are globally structurally identifiable. An identical conclusion can also be reached by considering the (linear and closed) expression for the mean of $\mathbf{x}(t)$, from which we recover an ODE model with coefficient matrix $\mathbf{A}$, and an analytical solution involving a sum of exponentials with exponents similarly involving the eigenvalues of $\mathbf{A}$. 
	
	To establish additional identifiable parameter combinations relating to higher-order statistical moments, we consider the case that $\mathbf{x}_1(0) = \mathbf{x}_0$ is fixed (else, as in \cref{fig1}, we can translate time). The appropriate initial condition for the unobserved states assumes that they follow the conditional stationary distribution	$\mathbf{x}_2(0) | (\mathbf{x}_1(0) = \mathbf{x}_0) \sim \mathcal{N}(\hat{\bm{\mu}}_2,\hat{\bm{\Sigma}}_{22})$ where $\hat{\bm{\mu}}$ and $\hat{\bm{\Sigma}}$ are parameter and initial condition dependent, and where $\hat{\bm{\mu}}_1 = \mathbf{x}_0$ and $\hat{\bm{\Sigma}}_{11} = \mathbf{0}$. Then,
		\begin{equation}
			\mathbf{x}(t) | (\mathbf{x}_1(0) = \mathbf{x}_0) \sim \mathcal{N}\Big((\mathbf{I} - \mathrm{e}^{-\mathbf{A} t}) \mathbf{b} + \mathrm{e}^{-\mathbf{A} t}\hat{\bm{\mu}}, \mathbf{\Sigma}(t)\Big),
		\end{equation}
	where
		\begin{equation}\label{eq:ndou_sigma}
			\mathrm{vec}(\bm\Sigma(t)) = (\mathbf{A} \oplus \mathbf{A})^{-1} \Big(\mathbf{I} - \mathrm{e}^{-(\mathbf{A} \oplus \mathbf{A})t}\Big) \mathrm{vec}(\mathbf{S}\mathbf{S}^\intercal) + \mathrm{vec}\Big(\mathrm{e}^{-\mathbf{A} t}\hat{\mathbf{\Sigma}}\mathrm{e}^{-\mathbf{A}^\intercal t}\Big),
		\end{equation}
	is the time-variant covariance function given the semi-fixed initial condition \cite{Meucci.2009}.
		
	Next, we consider that $\big[\bm\Sigma(t)\big]_{11}$ and its derivatives are observed. From \cref{eq:ndou_sigma}, it follows that
		\begin{equation*}
			\mathbf{\Sigma}'(0) = \mathbf{S}\mathbf{S}^\intercal + \mathbf{A} \hat{\bm{\Sigma}} + \hat{\bm{\Sigma}}\mathbf{A}^\intercal,
		\end{equation*}
	such that
		\begin{equation*}
			\big[\bm\Sigma(0)\big]_{11} = \big[\mathbf{S}\mathbf{S}^\intercal\big]_{11},
		\end{equation*}
	since $\hat{\bm{\Sigma}}_{11} = \mathbf{0}$ and so $[\mathbf{A} \hat{\bm{\Sigma}}]_{11} = [\hat{\bm{\Sigma}}\mathbf{A}^\intercal]_{11} =\mathbf{0}$. Therefore, we conclude that all parameter combinations that are elements of $[\mathbf{S}\mathbf{S}^\intercal]_{11}$ are identifiable (in the two-dimensional model considered previously, this corresponds to the $p^2$ term).
	
	We highlight that $[\mathbf{S}\mathbf{S}^\intercal]_{11}$ together with the eigenvalues of $\mathbf{A}$ do not form an exhaustive set of identifiable parameter combinations. In the two-dimensional case, for instance, the final combination $(c p - a r)^2 + a^2 s^2$ in fact comes from the necessary equivalence of the stationary variance $[\bm\Sigma_\infty]_{11}$.

	\subsection{Geometric noise}\label{sec:geo}
	
	We next consider a simple geometric extension to the partially observed Ornstein-Uhlenbeck process (\cref{sde1}) by considering that the noise magnitude scales with $\mathbf{x}$. The dynamics of the general $n$-dimensional process are given by
		\begin{equation}\label{sde2}
			\dd \mathbf{x}(t) = -\mathbf{A}\left(\mathbf{x}(t) - \mathbf{b}\right) \dd t + \mathrm{diag}(\mathbf{x}(t))\, \mathbf{S} \,\dd\mathbf{W},\\
		\end{equation}
	where $\mathbf{A}$, $\mathbf{b}$, and $\mathbf{S}$ are defined in \cref{linear_mat_pars}. We focus on the two-dimensional case where $\mathbf{x}(t) = [x(t),y(t)]^\intercal$ and only $x(t)$ is observed. While \cref{sde2} is still linear (such that the moments are closed at every order), analysis differs from the Ornstein-Uhlenbeck as \cref{sde2} may not possess an analytical solution and is not Gaussian: the system is no longer fully characterised by the system of moments up to finite order. 
	
	Applying It\^o's lemma to \cref{sde2}, we arrive at a recurrence-like relation that characterises the moment equations, given by
		\begin{equation}\label{sde2_recurrence}
		\begin{aligned}
			m_{i,j}'(t) &= \left(ij pr -ai - dj + \dfrac{i(i-1)p^2 + j(j-1)(r^2 + s^2)}{2} \right) m_{i,j}(t) \\
			&\qquad + \underbrace{(ae + bf) i m_{i-1,j}(t) + (ce + df) j m_{i,j-1}(t)}_{\text{Order }i+j-1}\\
			&\qquad - \underbrace{\Big(b i m_{i-1,j+1}(t) + c j m_{i+1,j-1}(t)\Big)}_{\text{Order }i+j}.\\
		\end{aligned}
		\end{equation}
	As the system is linear, it is closed at every order: the equation for $m_{i,j}'(t)$ depends only on moments of order $i + j$ and lower.
	
	\subsubsection*{Moment equation approach}
	
	The simplest way to assess identifiability of the geometric model is to apply established software directly to a finite system of moments \cite{Browning.2020,Dong.2023}. Considering the system of moments up to second order (both two-dimensional linear models considered thus far behave identically to first order), the set of identifiable parameter combinations is given by \textit{StructuralIdentifiability.jl} \cite{Dong.2023} as
		\begin{equation}\label{gb_ident}
			\left\{a,d,bc,bf,e,pr,r^2,s^2\right\}.
		\end{equation}
	Therefore, if without loss of generality we again assume that $p,s > 0$, then all parameters are identifiable up to a rescaling of the state variable $\hat{y}(t) = by(t)$. That is, all rescaled parameters in the scaled system governing $(x,\hat{y})$ are identifiable. Thus, we do not expect to gain any information about the parameters by considering moments of order three and higher.

	\subsubsection*{DA through recurrence relation}
	
	The second way that identifiability can be assessed is by ``projecting'' the recurrence relation (\cref{sde2_recurrence}) onto a relation involving higher order moments where $j = 0$ (i.e., that only include moments of the scalar observed variable $x(t)$).

	We can visualise the recurrence relationship through the stencil
	\begin{equation}
		\begin{pmatrix} 
			- 			& m_{i-1,j} & m_{i-1,j+1} \\
			m_{i,j-1} 	& m_{i,j} & -\\
			m_{i+1,j-1} & - & -
		\end{pmatrix},
	\end{equation}
	which we take to include moments that appear in the recurrence relation governing $m_{i,j}'(t)$. What is clear from the format of the stencil for the geometric model is that we can always solve \cref{sde2_recurrence} for $m_{i-1,j+1}$, and therefore all moments with $j = k + 1$, in terms of moments with $j \le k$ and their corresponding derivatives. Iterating this process, we see that it is theoretically possible to write all moments $m_{i,j}(t)$ with $j \ge 1$ as a linear combination of the observed moments $m_{i,0}(t)$ and their corresponding derivatives. An algorithm detailing this approach is given in \cref{app1}.

	First, consider the closed system of moments up to order one, dependent only on $m_{1,0}(t)$ and $m_{0,1}(t)$. From the relation for $m_{1,0}'(t)$ (i.e., \cref{sde2_recurrence} with $(i,j) = (1,0)$) we solve for $m_{0,1}(t)$ to obtain
		\begin{equation}
			m_{0,1}(t) = \dfrac{ae + bf - a m_{1,0}(t) - m_{1,0}'(t)}{b}.
		\end{equation}
	Differentiating and equating with the relation for $m_{0,1}'(t)$ yields a necessarily satisfied equation in terms of only the observed variables, given for $b \neq 0$ (otherwise, we lose the coupling between the states) by
		\begin{equation}\label{geo_io}
			0 = (bc - ad) e + (ad - bc) m_{1,0}(t) + (a + d) m_{1,0}'(t) + m_{1,0}''(t).
		\end{equation}
	Within the DA approach to structural identifiability, such a necessarily satisfied equation is typically referred to as an \textit{input-output} equation. Treating it as a monic polynomial in the directly observed states (and derivatives) $\{m_{1,0}(t),m_{1,0}'(t),m_{1,0}''(t)\}$, the structurally identifiable combinations are given by the coefficients, which uniquely define the polynomial and, by extension, the input-output relation. Therefore, from \cref{geo_io}, we see that the combinations $\{e, ad-bc, a + d\}$ are structurally identifiable. We may repeat the process at second order by solving for both $m_{1,1}(t)$ and $m_{0,2}(t)$ in terms of the observed states, and equating with the expression for $m_{0,2}'(t)$, to obtain a set of coefficients that can be reduced to the full set of identifiable combinations given in \cref{gb_ident} (supplementary code).

\section{Partially observed non-linear systems}

	In our previous work \cite{Browning.2020}, we approximate structural identifiability analysis for non-linear systems by generating a closed system of moment equations using an appropriate \textit{moment closure} technique \cite{Ruess:2013}. The resultant \textit{closed} system of ODEs can be processed using software, for instance \textit{StructuralIdentifiability.jl} \cite{Dong.2023}, that applies DA to produce the input-output equations from which structural identifiability can be assessed. There are currently no approaches, however, that can assess systems of ODEs that are not closed. In this section, we demonstrate how the recurrence-relation-based DA framework can be applied \textit{exactly} to a class of non-linear two-dimensional systems. In all cases, we consider a two-dimensional system in the states $\mathbf{x}(t) = [x(t),y(t)]^\intercal$ where only the state $x(t)$ is observed.

	\subsection{Semi-logistic model}\label{sec:semilogistic}

	The first non-linear model we consider retains linearity in the unobserved state $y(t)$. Specifically, we consider an extension of the two-dimensional Ornstein-Uhlenbeck model with a quadratic non-linearity in $x(t)$ through two logistic terms. We assume that the dynamics are now governed by
		\begin{equation}
		\begin{aligned}
			\dd x &= \big(ax (1 - b x) + c y\big) \dd t + p \,\dd W_1,\\
			\dd y &= \big(dx (1 - e x) + f y\big) \dd t + r \,\dd W_1 + s \, \dd W_2,
		\end{aligned}
		\end{equation}
	and that only $x(t)$ is observed. Setting $p = r = s = 0$ recovers a deterministic model where the identifiable parameter combinations are $\{ab,a+f,af-cd,abf-cde\}$ (i.e., where no individual parameters are identifiable). 
	
	The moments are defined by the recurrence relation
		\begin{equation}\label{semilogistic _recurrence}
		\begin{aligned}
			m_{i,j}'(t) &= (a i + f j) m_{i,j}(t) \\
			&\qquad + \underbrace{\dfrac{i(i-1)p^2m_{i-2,j}(t) + j(j-1)(r^2 + s^2)m_{i,j-2}(t)}{2} + ijpr m_{i-1,j-1}(t)}_{\text{Order }i+j-2}\\
			&\qquad + \underbrace{ci m_{i-1,j+1}(t) + d j m_{i+1,j-1}(t)}_{\text{Order }i+j}\\
			&\qquad - \underbrace{\Big(dej m_{i+2,j-1}(t) + a b i m_{i+1,j} \Big)}_{\text{Order }i+j+1},\\
		\end{aligned}
		\end{equation}
	which corresponds to the stencil
		\begin{equation}
		\begin{pmatrix}
			- & - & m_{i-2,j} & - \\
			- & m_{i-1,j-1} & - & m_{i-1,j+1}\\
			m_{i,j-2} & - & m_{i,j} & -\\
			- & m_{i+1,j-1} & m_{i+1,j} & - \\
			- & m_{i+2,j-1} & - & -
		\end{pmatrix}.
		\end{equation}

	Clearly, the system cannot be closed: the quadratic non-linearity introduces terms of order $i + j + 1$ to the equation governing the behaviour of moments of order $i + j$. However, we see that it is still possible to apply the DA-based procedure from the previous section to solve for moments with $j = k + 1$ first in terms of moments with $j = k$ and therefore, through iteration, in terms of the observed moments ($j = 0$) and their corresponding derivatives. Substituting into the governing equations for $m_{0,j}$ for $j \ge 1$ provides a set of necessarily satisfied equations that can be used to assess structural identifiability.
	
	To obtain the first equation, we consider the relation for $m_{1,0}'(t)$ and solve for $m_{0,1}(t)$. Differentiating and substituting into the governing equation for $m_{0,1}'(t)$, we obtain the necessarily satisifed equation
		\begin{equation}\label{nse_semilogistic}
			0 = m_{1,0}''(t) + a b m_{2,0}'(t) - (a + f) m_{1,0}'(t) + (c d e - a b f) m_{2,0}(t)  + (af - cd) m_{1,0}(t).
		\end{equation}
	As the necessarily satisfied equation is homogeneous, to make conclusions about structural identifiability, we must assume that a largest proper subset of monomial terms (for example, the terms $\{m_{1,0}(t),m_{2,0}(t),m_{1,0}'(t),m_{1,0}''(t)\}$) are linearly independent. We expect this to be the case, as a linear dependency would imply that a simpler necessarily satisfied equation can be derived. In general, we conjecture that, for systems fully characterised only by an infinite system of moments, the linear independence requirement will hold for all necessarily satisfied equations derived using our method (although we are not able to prove as such). From \cref{nse_semilogistic}, therefore, we see that the set of identifiable combinations is \textit{at least} the same as that for the corresponding ODE model. In fact, we expect this to be true for all polynomial SDEs that are linear in the unobserved variables (see \Cref{app2}). 

	Repeating the procedure for $m_{0,2}(t)$, we obtain the lengthy expression
		\begin{equation}\label{semilog_o2}
		\begin{aligned}
			0 &= 3 m_{2,0}'''(t) + 8ab m_{3,0}''(t) - 9(a + f) m_{2,0}''(t)+ 6 a^2 b^2 m_{4,0}'(t)
			\\&\qquad + \big(10 c d e - 2 a b (6 a + 11 f)\big) m_{3,0}'(t)+ 6\big(a^2 + 4 a f + f^2 - 2 c d \big) m_{2,0}'(t)
			\\&\qquad  - 6c^2 \left(m_{1,0}'(t)\right)^2 + 6\big(2 c f p - a b p^2\big) m_{1,0}'(t)+12\big(a b c d e - a^2 b^2 f\big) m_{4,0}(t)
			\\&\qquad  + 12\big(a b f (2 a + f) - c d (a (b + e) + e f)\big) m_{3,0}(t)-12 (a + f) (-c d + a f) \,m_{2,0}(t)
			\\&\qquad +12 (abf - c d e) p^2 m_{1,0}(t) - 6 f (a + f) p^2 + 6 c p (d p + 2 f r) - 6 c^2 (r^2 + s^2).
		\end{aligned}
		\end{equation}
	The set of additionally identifiable parameter combinations is again given by the coefficients. Assuming that $p > 0$ and that none of the earlier identifiable parameter combinations vanish, the set of additionally identifiable parameter combinations is at least given by $\{p,c^2,(fp - c r)^2 + c^2 s^2\}$. Interestingly, these combinations imply that if we assume that noise only enters the system through the observed variable (i.e., $r = s = 0$), then $f$ will become identifiable. This property is lost if $p = 0$, which will recover the deterministic model in which no individual parameters are identifiable.

	As the stochastic system is only fully characterised by the infinite system of moments, while our approach is exact, it is not exhaustive. Proceeding to third order, for example, may yield additional parameter combinations that are structurally identifiable. 	
	
	\subsection{Lotka-Volterra model}
	
	The next model at which we attempt our approach contains a non-linearity in both variables, in both equations. Specifically, we consider the Lotka-Volterra-like system subject to state-independent noise, governed by
	\begin{equation}
	\begin{aligned}
		\dd x &= (ax + b xy) \dd t + p \,\dd W_1,\\
		\dd y &= (cy + d xy) \dd t + s \,\dd W_2,
	\end{aligned}
	\end{equation}
	where only $x(t)$ is observed. In the ODE system recovered when $p = s = 0$, we have that the parameter set $\{a,c,d\}$ is structurally identifiable. 
	
	The moments are associated with a recurrence relation (given in full in the supplementary code) that corresponds to the stencil
	\begin{equation}
	\begin{pmatrix}
		- & - & m_{i-2,j} & - \\
		- & - & - & -\\
		m_{i,j-2} & - & m_{i,j} & m_{i,j+1} \\
		- & - & m_{i+1,j} & - 
	\end{pmatrix}.
	\end{equation}
	Fundamentally, the stencil for the Lotka-Volterra model differs from the semi-logistic model in that the higher order (i.e., $> i + j$) moments involve moments that are higher order in the unobserved variable (e.g., $m_{i,j+1}$). We can still solve for at least moments with $j = 1$; specifically, all moments of the form $m_{1,j}$, in terms of observed moments $m_{i,0}$ and their derivatives. However, we get no information about $m_{0,j}$ (by definition, $m_{0,0}(t) \equiv 1$) and hence cannot form necessarily satisfied equations from which we can evaluate structural identifiability.
	
	\subsection{Simplified Lotka-Volterra model}
	
	Next, we consider a simplified Lotka-Volterra model with the same non-linear governing equation for the unobserved variable, but with a linear governing equation in the observed variable. Specifically, we consider the model
	\begin{equation}\label{lv_simple}
	\begin{aligned}
		\dd x &= (ax + b y) \dd t + p \,\dd W_1,\\
		\dd y &= (cy + d xy) \dd t + s \,\dd W_2,
	\end{aligned}
	\end{equation}
	where only $x(t)$ is observed.	Again, the ODE system recovered when $p = s = 0$ yields the parameter set $\{a,c,d\}$ as structurally identifiable.  
	
	The moments are associated with a recurrence relation that corresponds to the stencil
	\begin{equation}
	\begin{pmatrix}
		- & - & m_{i-2,j} & - \\
		- & - & - & m_{i-1,j+1}\\
		m_{i,j-2} & - & m_{i,j} & - \\
		- & - & m_{i+1,j} & - 
	\end{pmatrix}.
	\end{equation}
	Again, the system cannot be closed. However, for the simplified Lotka-Volterra model we can solve for moments with $j = k + 1$ in terms of moments and derivatives with $j \le k$ and, by iteration, the observed moments with $j = 0$. We can then proceed as in \cref{sec:semilogistic} by equating with the governing equations for $m_{0,j}'(t)$ to iteratively form a set of necessarily satisfied equations that give information about structural identifiability.

	At first order, the necessarily satisfied equation is given by
		\begin{equation}
			0 = d p^2 +2 ac m_{1,0}(t) + 2ad m_{2,0}(t) - 2(a + c)m_{1,0}'(t) - d m_{2,0}'(t) + 2 m_{1,0}''(t),	
		\end{equation}
	such that the set of identifiable parameter combinations is given by (at least) $\{a,c,d,p^2\}$. Progressing to second order provides no new information, and progressing through to order three indicates that $b^2 s^2$ is also identifiable (supplementary code). We expect this to be exhaustive through the rescaling $b y \mapsto \hat{y}$, so we conclude that the system is structurally identifiable, up to the scaling of the unobserved variable $y$. 
		
	\subsection{Chemical Langevin equation}
	
	Finally, we consider the structural identifiability of the second-order chemical reaction network
	\begin{equation}\label{crn}
		X + X \underset{\beta}{\stackrel{\alpha}{\rightleftharpoons}} Y,\qquad \varnothing \underset{\delta}{\stackrel{\gamma}{\rightleftharpoons}} Y, \qquad \varnothing \underset{\zeta}{\stackrel{\varepsilon}{\rightleftharpoons}} X,
	\end{equation}
	in the case that only $x(t)$, corresponding to the concentration of the molecule $X$, is observed. Here, all rate parameters are assumed to be non-negative. The corresponding SDE system that we analyse derives from a chemical Langevin equation approximation, given by 
	\begin{equation}\label{sde_cle}
	\begin{aligned}
		\dd \begin{pmatrix} x\\y\end{pmatrix} &= %
			\begin{pmatrix} 
				-x (2x\alpha + \zeta) + 2\beta y + \varepsilon \\ 
				\alpha x^2 - (\beta + \delta) y + \gamma
			\end{pmatrix} \dd t \\&\qquad+ %
			\begin{pmatrix}
				-2 \sqrt{\alpha} x & 2\sqrt{\beta y} & 0 & 0 & \sqrt\varepsilon & \sqrt{\zeta} x\\ 
				\sqrt{\alpha} x & - \sqrt{\beta y} & \sqrt\gamma & -\sqrt{\delta y} & 0 & 0
			\end{pmatrix} \dd \mathbf{W}_t,
	\end{aligned}
	\end{equation}
	where $\mathbf{W}_t \in \mathbb{R}^6$ is a six-dimensional Wiener process. Compared with previous models analysed in this section, \cref{sde_cle} contains non-linearities in both the drift term and the diffusion term. 
	
	Identifiability of chemical reaction networks in the \textit{small} molecule limit (i.e., modelled as a Poisson process) has previously been explored in \cite{Encisco.2021}, albeit not for the partially observed non-linear system given in \cref{crn}. In the large molecule limit, we recover an ODE system governed by the drift of \cref{sde_cle} with identifiable parameter combinations given by $\{\alpha,\delta,\beta+\zeta,(\beta + \delta)\zeta,2\beta\gamma + (\beta + \delta)\varepsilon\}$. Here, we see that $\beta$ and $\zeta$ are at least \textit{locally identifiable}: since the parameter combinations are quadratic, there are potentially multiple disconnected parameter sets that preserve the identifiable parameter combinations. In the case that one root root yields a biologically infeasible negative growth rate constant, these parameters are potentially globally identifiable. The SDE system, on the other hand, corresponds to an intermediate molecule count regime. As we can recover the ODE system through a limit and a rescaling, we expect that the structurally identifiable combinations in the SDE model correspond \textit{at least} to those that exist for the ODE model. Furthermore, we highlight that the chemical Langevin equation derived model is the only model we consider where the diffusion term relates to process noise. Consequentially, moving from the ODE to the SDE model does not introduce additional parameters.

	The moments of the SDE system (\cref{sde_cle}) are associated with a recurrence relation that corresponds to the stencil
		\begin{equation}
		\begin{pmatrix}
			- & - & m_{i-2,j} & m_{i-2,j+1} \\
			- & - & m_{i-1,j} & m_{i-1,j+1}\\
			m_{i,j-2} & m_{i,j-1} & m_{i,j} & - \\
			- & m_{i+1,j-1} & m_{i+1,j} & - \\
			m_{i+2,j-2} & m_{i+2,j-1} & - & - 
		\end{pmatrix},
		\end{equation}
	which, while involving terms of order $i + j + 1$, is similar in form to the stencils analysed previously. Therefore, our approach can be applied directly, with the caveat that solving $m_{i,j}'(t)$ for $m_{i-1,j+1}(t)$ will involve the term $m_{i-2,j+1}(t)$; thus, we must proceed iteratively in our formulation of expressions for the unobserved moments in terms of the observed moments and their derivatives.
		
	At first order, we obtain the necessarily satisfied condition
		\begin{equation}
			0 = m_{1,0}''(t) + 2\alpha m_{2,0}'(t) + (\beta + \delta + \zeta)m_{1,0}'(t) + 2\alpha\delta m_{2,0}(t) + (\beta + \delta)\zeta m_{1,0} - (\beta + \delta)\varepsilon - 2\beta\gamma,
		\end{equation}
	which gives the same set of identifiable parameter combinations as the corresponding ODE model (the large molecule limit), previously given. Applying the same procedure to second-order yields a lengthy necessarily satisfied equation (supplementary code), which, following reduction, indicates that the combination $4\varepsilon + 3\zeta$ is structurally identifiable. As we have now identified six irreducible structurally identifiable combinations of the six parameters, and can show that the the map from parameters to identifiable combinations is one-to-one (through nonsingularity of the Jacobian, see supplementary code), we conclude that \textit{all} parameters are at least locally identifiable (again, since the identifiable combinations are quadratic).

\section{Discussion}

Stochasticity is fundamental to many biological processes, necessitating the application of SDEs or other stochastic models in the interpretation of associated data. Even for systems that are well characterised by deterministic models, calibrating a stochastic counterpart to resultant data can often extract additional information: in some cases rendering non-identifiable parameters identifiable \cite{Browning.2020}. Establishing the structural identifiability of model parameters is an essential prerequisite for model choice, experimental design, and reparameterisation \cite{Raue:2009,Chis:2011b,Eisenberg:2014}. For stochastic models, in particular, inference methods that operate on partially observed sample paths are computationally costly \cite{Andrieu.2010, Warne:2020}. Any method able to establish identifiable parameter combinations is likely to be of value, and potentially able to accelerate inference by predicting the shape of the resultant posterior distribution (or likelihood surface). Yet, to the best of our knowledge, there exists no general methodology that can assess structural identifiability of stochastic models. In this work, we extend the DA approach to partially observed SDE models to present a moment-based framework for exact structural identifiability analysis of both linear and a class of two-dimensional non-linear models.

For linear models, a finite closed system of moment equations can be derived at every order, allowing for direct assessment of identifiability using existing methods \cite{Browning.2020}. For linear models with purely additive (i.e., constant) noise, system behaviour is entirely characterised by a second-order system of moments, thus the resultant identifiability conclusions are also complete. For general linear systems, this does not hold: a fundamental limitation of our approach is that our analysis is not exhaustive for both general linear and non-linear systems. That is, we cannot guarantee that additional identifiable parameter combinations will not be found through analysis of higher-order necessarily satisfied equations. Exacerbating this challenge is the rate at which complexity increases in the necessarily satisfied equations, even for relatively simple models (e.g., \cref{semilog_o2}). While iteratively constructing these equations is relatively straightforward (\cref{app1}), distilling coefficients into an irreducible set of identifiable parameter combinations is not; although, more sophisticated computer algebra techniques can potentially address this bottleneck \cite{Ovchinnikov.2023}. Alternatively, future work may focus on methodology to derive from the recurrence relation a general necessarily satisfied equation from which identifiability can be assessed directly.

A question most pertinent to linear models, where the analogous ODE model corresponds exactly to the mean of the stochastic process, is whether identifiability of the stochastic model is improved relative to that of the deterministic model. We observe in \cref{sec:semilogistic}, and establish generally in \cref{app2}, that SDE models that are linear in the unobserved variable are at least as identifiable as the associated ODE model (that is, the identifiable parameter combinations of the ODE model are a subset of that of the SDE). Specific results show that the geometric linear model is structurally identifiable (up to a rescaling); an improvement on the analogous ODE model, in which no individual parameters in the coefficient matrix $\mathbf{A}$ are identifiable. Findings for the additive linear model are less clear (\cref{sde1_si}). While the number of structurally identifiable parameter combinations increases from three to five, the SDE introduces three additional parameters (this increase in parameter count is not the case for models derived from the chemical Langevin equation, for example). Only by imposing additional knowledge about the model structure do we gain a clear improvement to model identifiability. Specifically, if we assume that the noise process only affects the observed variable (i.e., $p \neq 0$, $r = s = 0$), all parameters in $\mathbf{A}$ become identifiable (up to a rescaling of the units of $y(t)$). What remains unclear is how theoretical improvements in structural identifiability translate practically to uncertainty in parameter estimates. For instance, a parameter may become structurally identifiable following analysis of a higher order necessarily satisfied equation, while practical uncertainty in observations of higher order moments may still render the parameter practically non-identifiable. Regardless, better capturing observed variability with intrinsic stochasticity may improve the accuracy of parameter estimates, even in cases where additionally introduced noise parameters remain non-identifiable \cite{Browning.2022}.

\begin{figure}
	\centering
	\includegraphics{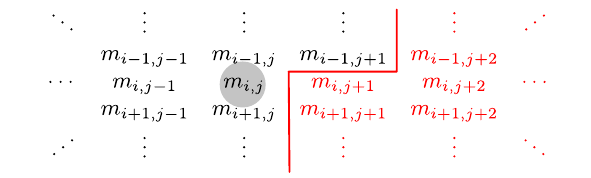}
	\caption{\textbf{Model applicability.} Our method can analyse the structural identifiability of two-dimensional SDE models where the governing equation for $m_{i,j}'(t)$ only depends on moments $m_{p,q}(t)$ that satisfy $q \le j$ or, for $q = j + 1$, $p < i$ (allowed terms in black, disallowed terms in red).}
	\label{fig4}
\end{figure}

Most novel about our new approach is that it can establish structurally identifiable parameter combinations of a broad class of non-linear models without resorting to a moment closure approximation. While there are likely some exceptions, this class includes all two-dimensional polynomial SDE models that are linear in the unobserved variable, and non-linear models where the governing equation for $m_{i,j}$ is \textit{not} dependent on higher order moments that are higher order in the observed variable (that is, dependent only on moments $m_{p,q}(t)$ that satisfy either $q \le j$, or $q = j + 1$ and $p < i$; see \cref{fig4}). This ``delayed dependence'' on higher-order moments in the unobserved variable $j > 0$ allows us to both recursively solve for the unobserved moments in terms of the observed moments and, crucially, construct a necessarily satisfied equation. These restrictions arise primarily from the linearity of the (infinite) system of moment equations, which limits the information obtainable through repeated differentiation (DA-based approaches for ODE models, for instance, rely on an expanded number of satisfied equations obtained through differentiation). Consequentially, we do not expect our approach to scale well to higher dimensional systems with more than one unobserved variable due to the comparatively higher number of unobserved moments that need to be eliminated. It remains unclear whether it is possible to apply DA- and moment-based methods to more general non-linear systems. Alternative approaches could include those based on expansions, such as polynomial chaos \cite{Xiu:2002}, or to perform analysis directly on the infinite linear system of moments using structural identifiability tools based on Taylor series, Lie algebra, or Laplace transforms \cite{Chis:2011}.

For ODE models, structural identifiability typically corresponds to a hypothetical regime comprising infinite, noise-free data: in some cases from a single trajectory initiated with a single initial condition. For SDE models, our approach is to define structural identifiability in a distributional sense  (\cref{def_sde}), effectively considering an infinite quantity of noiseless data such that the entire distribution of model outputs is itself observed. If the process is at statistical equilibrium, this data may comprise a single trajectory, observed for an infinite duration of time. 

Consideration of the initial condition differs, therefore, to typical identifiability analysis of deterministic models, in which the initial condition is a fixed quantity and able to be treated as a parameter. For stochastic models, the initial condition can always be interpreted as a (potentially degenerate) probability distribution, parametrisable by the infinite system of moments. As we demonstrate comprehensively for linear models in \cref{seclinear}, the default initial condition implicitly assumed is stationarity. Imposing other initial conditions can only ever yield a larger number of structurally identifiable parameter combinations, due to the natural fact that all systems are eventually (in the infinite-data regime) observed at stationarity. For the linear model, the most information arises out of the perturbed initial condition, under which all parameters are identifiable. An intuitive reason for this is that the perturbed initial condition is analogous to an ODE model with known external pulse at $t = 0$. It remains unclear how such an input to SDE models can be considered explicitly, although this more explicit input-orientated approach may allow for analysis of perturbed non-linear models. Further complicating the analysis of non-linear models (and unconstrained linear models) is that we are no longer guaranteed that a stationary solution exists. As our method draws conclusions, effectively, from the interrelationships within the temporal dynamics of observed moments, we still expect the structurally identifiable parameter combinations to be valid. For models in which a non-degenerate stationary solution does not exist (for example, models where states either tend to zero or infinity), an experimentally relevant initial condition must be considered explicitly.

Establishing structural identifiability is important for model parameterisation and, crucially, model choice. Despite this, tools that enable such analysis of models that are not ODEs remain in their infancy. In this work, we present methodology that can establish structurally identifiable parameter combinations in a large class of partially observed linear and non-linear SDE models. While our focus is on two-dimensional SDE models, our approach may translate to related systems: for example, partially observed partial differential equation models through the Fokker-Planck description of the SDE. Most importantly, however, we define structural identifiability for stochastic models and lay the foundation for future work in this as yet understudied area.

\section*{Data availability}
	Mathematica and Julia code used to perform the symbolic and numerical computations, respectively, available on GitHub at \url{https://github.com/ap-browning/sde_structural_identifiability}.

\section*{Acknowledgements}
	APB thanks the Mathematical Institute for a Hooke Fellowship. REB and TEL are supported by a grant from the Simons Foundation (MP-SIP-00001828). HR acknowledges funding from the Royal Society grants UF150238 and URF/R/211032. For the purpose of open access, the author has applied a CC BY public copyright licence to any author accepted manuscript arising from this submission.


\appendix
\setcounter{equation}{0}\renewcommand\theequation{A\arabic{equation}}
\section{Appendix}

\subsection{Algorithm of implementation}\label{app1}

\begin{algorithm}[H]
\caption{}
\label{alg1}
\begin{algorithmic}[1]
	\item Given a two-dimensional system of stochastic differential equations of the form
		\begin{equation}
		\begin{aligned}
			\dd \mathbf{x} &= \mathbf{f}(\mathbf{x};\bm\theta) \,\dd t + \mathbf{g}(\mathbf{x};\bm\theta) \,\dd \mathbf{W},\\
		\end{aligned}
		\end{equation}
		for $\mathbf{x}(t) = [x(t),y(t)]^\intercal$, where $x(t)$ is the observed variable and $y(t)$ is the unobserved variable. Here, $\mathbf{f} : \mathbb{R}^2 \mapsto \mathbb{R}^2$, $\mathbf{g} : \mathbb{R}^2 \mapsto \mathbb{R}^{2 \times m}$, with both $\mathbf{f}$ and $\mathbf{g}$ polynomial, $\bm\theta$ is a vector of parameters, and $\mathbf{W}$ an $m$-dimensional Wiener process.
	\item Compute the ordinary differential equation governing the dynamics of $m_{i,j}(t) = \langle x^i y^j \rangle$ for $i,j \in \mathbb{N} \cup \{0\}$ using It\^{o}'s lemma as follows.
			\begin{equation}\label{alg_ito}
				m_{i,j}'(t) = \left\langle \nabla_\mathbf{x} \left(x^iy^j\right) \cdot \mathbf{f}(\mathbf{x}) + \frac{1}{2}\mathrm{Tr}\left(\mathbf{g}^\intercal(\mathbf{x}) H_\mathbf{x}\left(x^iy^j\right)\mathbf{g}(\mathbf{x})\right) \right\rangle,
			\end{equation}
		where $\nabla_\mathbf{x}$ and $H_\mathbf{x}$ denote the gradient and Hessian with respect to $\mathbf{x}$, and $\mathrm{Tr}(\cdot)$ is the trace.
	\item As \cref{alg_ito} is polynomial, carry through the expectation and substitute $\left\langle x^p y^q \right\rangle \rightarrow m_{p,q}$. Verify that the method can be applied: all terms $m_{p,q}(t)$ should satisfy either $q \le j$ or $q = j + 1$ and $p < i$. Otherwise, the method cannot be applied. The result of this step will be a recurrence relation of the form 
			\begin{equation}\label{alg_rec}
				m_{i,j}'(t) = c_{0,0} + \sum_{p,q} c_{p,q} m_{p,q}(t).
			\end{equation}
	\item Formulate the first necessarily satisfied equation:
	\begin{algorithmic}[leftmargin=*]
		\item[\footnotesize 4.1:] From \cref{alg_rec}, form the governing equation for $m_{1,0}'(t)$ and solve for $m_{0,1}(t)$. 
		\item[\footnotesize 4.2:] Substitute the result of Step 4.1 into the governing equation for $m_{0,1}'(t)$, formulated using \cref{alg_rec}. The result will be an expression that only depends on moments $m_{p,q}$ with $q = 0$ and their respective derivatives.
	\end{algorithmic}
	\item[\footnotesize 5:] Establish structurally identifiable parameter combinations from the necessarily satisfied equation:
	\begin{algorithmic}[leftmargin=*]
		\item[\footnotesize 5.1:] Divide the necessarily satisfied equation by the coefficient of a term (for example, the coefficient of $m_{1,0}$) to ensure that the necessarily satisfied equation is monic.
		\item[\footnotesize 5.2:] The coefficients of the result of Step 5.1 are the structurally identifiable parameter combinations. For example, if the necessarily satisfied equation is
			\begin{equation*}
				0 = a + b + (a - b)m_{1,0}(t) + m_{1,0}'(t),
			\end{equation*} 
			then the set of structurally identifiable parameter combinations is $\{a+b,a-b\}$. 
		\item[\footnotesize 5.3:] (Optional) Reduce the set of structurally identifiable combinations by taking algebraic combinations of the coefficients. For the example given in the previous step, $c_1 = a + b$ and $c_2 = a - b$ are structurally identifiable. Therefore, $(c_1 + c_2) / 2 = a$ is identifiable, as is $(c_1 - c_2) / 2 = b$. If all model parameters are structurally identifiable, then the model is structurally identifiable. 
	\end{algorithmic}
	\item[\footnotesize 6:] Formulate and analyse higher-order necessarily satisfied equations:
	\begin{algorithmic}[leftmargin=*]
		\item[\footnotesize 6.1:] From \cref{alg_rec}, form the governing equation for $m_{0,j}'(t)$. This will result in a $j$th order necessarily satisfied equation. 
		\item[\footnotesize 6.2:] Eliminate terms $m_{p,q}(t)$ with $q > 0$ recursively as follows (proceed with $q$ largest to smallest, within a value of $q$ for $p$ largest to smallest). This can be done by forming the governing equation for $m_{p+1,q-1}'(t)$ (\cref{alg_rec}), solving for $m_{p,q}(t)$, and substituting the result into the equation for $m_{0,j}'(t)$. Each substituting will introduce additional ``lower order'' terms that will themselves be eventually be eliminated. 
		\item[\footnotesize 6.3:] Apply the procedure in Step 5 to establish additional structurally identifiable parameter combinations from the higher-order necessarily satisfied equation. Optional Step 5.3 can be applied for each additional higher-order necessarily satisfied equation analysed to obtain a reduced set of identifiable parameter combinations.
	\end{algorithmic}
\end{algorithmic}
\end{algorithm}
\vfill

\subsection{Identifiability of SDE and ODE models that are linear in the unobserved variable}\label{app2}

Consider partially observed polynomial systems of the form 
	\begin{equation}
	\begin{aligned}
	 	\dd x &= \left(\sum_{i=0}^n c_i x^i + a y\right)\dd t + p\,\dd W_1,\\
	 	\dd y &= \left(\sum_{i=0}^n d_i x^i + b y\right)\dd t + r\,\dd W_1 + s\,\dd W_2,
	\end{aligned}
	\end{equation}
for sufficiently large $n$. 

We can obtain a leading order input-output relation by taking expectations and eliminating moments of $y$ directly. We obtain after some simplification
	\begin{equation}
		0 = \langle x\rangle '' - (b + c_1) \langle x \rangle ' - \sum_{j=2}^n c_j \langle x^j \rangle' - \sum_{j=0}^n (a d_j + b c_j) \langle x^j \rangle,
	\end{equation}
giving the set of identifiable parameter combinations as $\{b + c_1\} \cup \{c_j\}_{j=2}^n \cup \{a d_j + b c_j\}_{j=0}^n$. 

For the equivalent ODE system, we note that $\langle x^i \rangle = \langle x \rangle^i$, and thus obtain
	\begin{equation}
		0 = \langle x\rangle '' - (b + c_1) \langle x \rangle ' - \sum_{j=2}^n  c_j j \langle x \rangle ' \langle x \rangle^{j-1} - \sum_{j=0}^n (a d_j + b c_j) \langle x^j \rangle,
	\end{equation}
leading to the same set of identifiable parameter combinations. 


\end{document}